\documentclass[fleqn,12pt,twoside]{article}
\usepackage{espcrc1}
\usepackage{psfrag}
\input epsf
\usepackage{graphicx}
\usepackage[figuresright]{rotating}

\newcommand{\AmS}{{\protect\the\textfont2
  A\kern-.1667em\lower.5ex\hbox{M}\kern-.125emS}}

\title{
       \vspace{-2.0cm}
       {\normalsize  ITEP-LAT/2002-30}     \\[-0.2cm]
       {\normalsize KANAZAWA 02-38}   \\[0.850cm]
Flux Tubes of Two- and Three-Quark System in Full
QCD\footnote{Talk given by H. Ichie}}

\author{
H. Ichie
\address{ 
Institute for Theoretical Physics, Kanazawa University, Kanazawa
920-1192, Japan},
        V.~Bornyakov$^{\rm a}$, 
        T.~Streuer\address{
        NIC/DESY Zeuthen, Platanenallee 6, D-15738 Zeuthen, Germany}
        and 
        G.~Schierholz$^{\rm b}$}
       
\begin{document}

% typeset front matter
\maketitle

\begin{abstract}

We study the abelian color flux of two- and three-quark systems
in the maximally abelian gauge in lattice QCD with dynamical fermions.
We find that the abelian flux tube formed between quark and antiquark 
is very much the same as in quenched QCD up to quark separations of
$R$$\sim$ 2fm. The profile of the color electric field 
in three-quark system suggests $Y$ ansatz, which might be 
interpreted as the result 
of the vacuum pressure in the confined phase. In order to clarify 
the flux structure, we investigate the color electric field 
of the three-quark system splittting the abelian gauge field into
the monopole and photon parts.

\end{abstract}

\section{Introduction}

% The chromoelectric flux tube (color flux tube) between static color 
%electric charges forms an interface between the superconducting 
%confining phase and the conducting phase, and in this capacity bears 
%valuable information about the properties of the QCD vacuum. 

In the dual superconductor picture of confinement, 
%proposed by 't Hooft, Nambu and Mandelstam more than 25 years ago,
the color flux tube is formed due to the
dual Meissner effect caused by monopole condensation \cite{nambu}.
%, which is dual structure of the 
%as the dual version of the Abrikosov vortex in the superconductor.
Such a picture is based on the abelian gauge theory 
obeying Maxwell's equations, 
and 't Hooft built a bridge between the abelian gauge theory and QCD 
introducing abelian gauge fixing and abelian projection \cite{nambu}.
In the last decade, lattice simulations, especially in the
maximally abelian (MA) gauge, presented evidence in favor of this scenario.
%of  
%the abelian dominance and  monopole condensation, 
%and also shows that the profile of the color electric field and 
%super monopole current are suggested by dual Amp\`ere law\cite{bss}. 
%The dual superconductor picture is considered to be one of the 
%plausible explanation for confinement or to include the relevant 
%degrees of the freedom for confinement mechanism.

The most previous lattice studies, however, have been performed in SU(2) 
gauge theory without dynamical quarks. We consider the more realistic
case of lattice QCD with dynamical quarks.  
We extend the study of the abelian color flux tube between static quark
and antiquark (Q$\bar{\rm Q}$) to SU(3) gauge group and make first investigation of the 
%the flux structure in more realistic case, namely, in full SU(3).
abelian color flux tube in the three-quark (3Q) system. 
%First, we will discuss whether an explanation with dual superconductor 
%scenario for the confinement is available also to SU(3) case.   
%Different from SU(2), where the meson and baryon are not different, 
%in SU(3) the baryonic structure can be discussible in addition to
%mesonic flux tube. Although there are many studies on
%mesonic flux tube, there is almost no investigation on baryonic flux
%using lattice simulation so far. Using the advantage of maximally
%abelian gauge, we will investigate the profile of the baryonic flux. 
%Of further interest is the effect of  dynamical quarks
%on 
%the flux tube, especially the string breaking problem\cite{schi}. 
In QCD the string 
formed between static quark and anti-quark breaks and two static-light mesons
\begin{figure}[th]
\vspace{-0.9cm}
%\vspace*{-0cm}
%\vspace{3cm}
%\begin{flushright}
\begin{center}
%\vspace{3cm}
\epsfxsize=8.7cm
\epsfbox{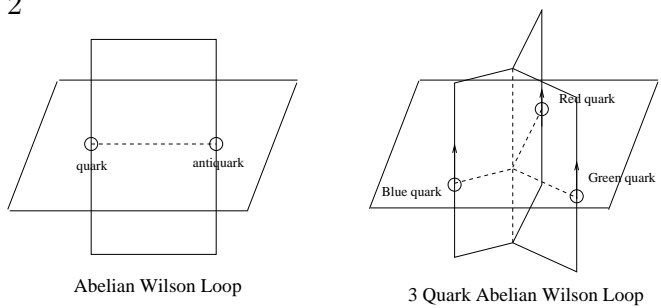}
\vspace{-1cm}
%\end{flushright}
\caption{Abelian Wilson loop for 2-quark system(left) and 3-quark
system(right). }
\end{center}
\vspace{-0.6cm}
\end{figure}
are created, when the separation is large enough. 
We measure abelian flux tube with large quark separation and 
look for effects of the string breaking.

\section{Operators and simulation details}
 The observables describing local structure  of the flux tube
are determined from the correlation function between an appropriate operator
$O(s)$ and abelian Wilson loop \cite{bss,dikk}: 
%Local quantities of an abelian flux tube are obtained 
%by expectation value in the presence of quark charges:
%\begin{eqnarray}
%O(x) = \int D A_\mu O(x) {\rm exp}(iS + i {\vec j}_\mu \cdot {\vec A_\mu})  
%O(x) = \int D A_\mu O(x) {\rm exp}(iS + i {j}_\mu \cdot { A_\mu})  
%\nonumber 
%\end{eqnarray}
%\begin{eqnarray}
%{\vec j}_\mu =  \delta^{\mu 0} \{ \vec Q_{\rm R} \delta^3 (x-x_{\rm R})
%+ \vec Q_{\rm G} \delta^3 (x-x_{\rm G})
%+ \vec Q_{\rm B} \delta^3 (x-x_{\rm B}) \} \nonumber
%\end{eqnarray}
%where $\vec Q_{a} \equiv e \vec{a}_\alpha$ with  $\vec{a}_\alpha$
%weight vectors of SU(3) algebra.  On lattice, it corresponds
%by the correlation between an appropriate operator and abelian Wilson
%loop \cite{dikk}:
\begin{eqnarray}
\hat{O}(s) = \frac{\langle O(s) W_{\rm abel} \rangle}{\langle W_{\rm abel} \rangle }
 - \langle O \rangle.  
%O(s) = \frac{\langle O(s) W_{\rm 3Q} \rangle}{\langle W_{\rm 3Q} \rangle }
% - \langle O \rangle,  \nonumber
\end{eqnarray}
%3 quark Wilson loop 
%\begin{eqnarray}
% W_{\rm 3Q} = \frac{1}{3!} \epsilon_{abc} \epsilon_{a'b'c'}
% U^{aa'} \cdot U^{bb'} \cdot U^{cc'} \nonumber
%\end{eqnarray}
%\item
%where for C-parity even operator 
The abelian Wilson loop for Q$\bar{\rm Q}$ system is
defined using abelian link variables $u_\mu(s)$ as 
\begin{equation}
W_{\rm abel}^{{\rm Q}\bar{\rm Q}}(R,T) = \frac{1}{3}\,\mbox{Tr}\, \prod_{s \in {\mathcal C}} u_\mu(s)
=\frac{1}{3}\,\mbox{Tr}\,e^{{\rm i} \theta_{\mathcal C}},
\end{equation}
where ${\mathcal C}$ is a rectangular loop of extension $R \times T$,
%$\theta_{\mathcal C} = 
%diag\{\theta_{1,\mathcal C},\theta_{2,\mathcal C}, \theta_{3,\mathcal C}\}$
and  for 3Q system as \cite{dikk2} 
\begin{eqnarray}
 W^{\rm 3Q}_{\rm abel}= \frac{1}{3!} |\epsilon_{abc}|
 u^a_{\rm R} \cdot u^b_{\rm G} \cdot u^c_{\rm B}, \nonumber
 \hspace*{1.5cm} \mbox{with}
\hspace*{0.5cm} u^a_{ \mathcal C} = \prod_{s \in \Gamma_{\mathcal C} } u^a_\mu(s), 
\hspace*{1cm}
\mbox{ $\Gamma_{\mathcal C}= \Gamma_{\rm R},\Gamma_{\rm G},\Gamma_{\rm B}$}
\nonumber
\end{eqnarray}
%for C-parity odd operator 
%\begin{eqnarray}
% W_{\rm 3Q} = u^1_{\rm R} \cdot u^2_{\rm G} \cdot u^3_{\rm B} \nonumber
%\end{eqnarray}
with a path product 
%$u^a_{ \mathcal C}$      
%\begin{eqnarray}
%u^a_{ \mathcal C} = \prod_{s \in \Gamma_{\mathcal C} } u^a_\mu(s) \nonumber 
%\hspace*{1cm}
%\mbox{\large $\Gamma_{\mathcal C}= \Gamma_{\rm R},\Gamma_{\rm G},\Gamma_{\rm B}$}
%\nonumber
%\end{eqnarray}
of abelian link variables along the stapler-type path 
$\Gamma_{\mathcal C}$ shown in Fig.1. 
Unlike in the nonabelian Wilson loop case,
the color of quarks does not change during the propagation, 
because the off-diagonal components of the gauge field are frozen in the
abelian projected theory. Most measurements are done in the plane which
is a central time slice of the abelian Wilson loop shown also in Fig. 1.

The simulations were performed using lattice configurations 
generated with $N_f$=2 non-perturbatively O(a) improved Wilson fermions 
with 
%$\beta$=5.2, $\kappa$=0.1355 on $16^3\cdot 32$ lattices  and with 
at $m_\pi/m_\rho\sim 0.7$ \cite{booth}.  
%$\beta$=5.29, $\kappa$=0.1355($m_\pi/m_\rho\sim 0.7$) on
%$24^3\cdot 48$ lattices.  
The link variables are fixed to
maximally abelian gauge with a simulated annealing algorithm. % \cite{}. 
%The 3 static charges, red,
%green and blue quarks, are located at $(17,14)$ $(22,6)$ and $(12,6)$
%on $x$-$y$ plane, respectively.
For noise reduction, we used a smearing method for spatial links of 
the abelian Wilson loop. 

%\begin{figure}[b!t]
\begin{figure}[t]
\psfrag{Action Density}{{ Action Density}}
\psfrag{Monopole Density}{{ Monopole Density}}
\psfrag{Monopole Current}{{ Monopole Current}}
\psfrag{Color Electric Field}{ Color Electric Field}
\psfrag{action}{$\rho \cdot r^4_0 \beta$}
\psfrag{mono}{$\rho_k \cdot r^3_0 $}
\vspace{-2cm}
\hspace{-.0cm}\includegraphics[width=9.5cm,angle=270]{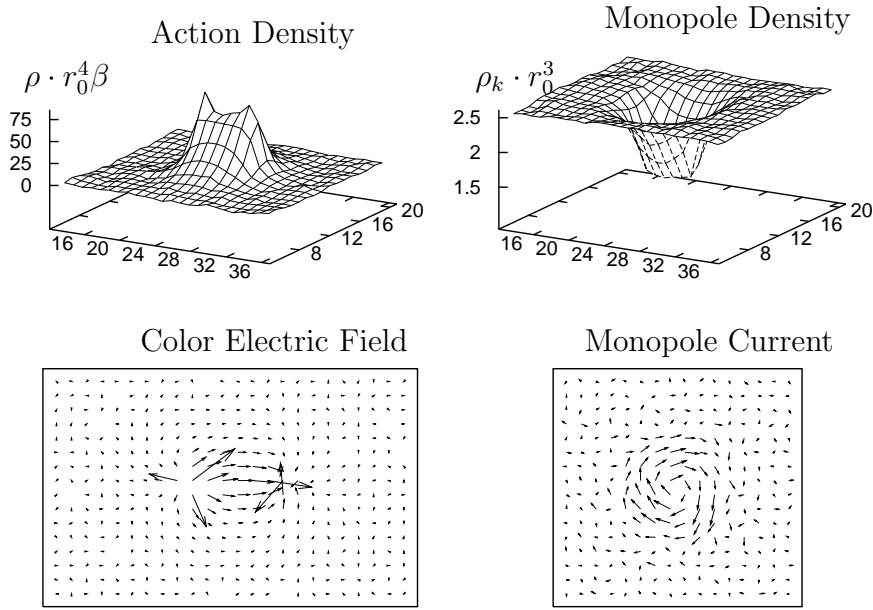}
\vspace{-.9cm}
\caption{\it Color flux tube in full QCD}
\vspace{-.4cm}
\end{figure}

\section{Abelian Flux Tube}

Fig.~2 shows the structure of Q$\bar{\rm Q}$  abelian flux tube 
in full lattice QCD for $R\sim$0.5 fm.
%We took $R=6$, which corresponds $R\sim$0.5 fm, and $T=6$ for the abelian
%Wilson loop $W_{\rm abel}^{{\rm Q}\bar{\rm Q}}(R,T)$. 
Similar to SU(2) \cite{bss} and SU(3) \cite{dikk2} gluodynamics, 
we observe enhancement  in the action density.
% and color electric flow from antiquark to quark in the electric field. 
Opposite to this, the monopole density is suppressed in agreement with vanishing
of the dual Higgs field in the center region of the flux tube.
Furthermore, in the plane perpendicular to the flux tube, the 
solenoidal monopole current is clearly seen.
%,  which constricts the color electric fields into flux tubes
%(This is expressed by the dual Amp\`ere law.), is found. 
All these observations are in accordance with the dual superconductor 
picture of confinement in particular with the assumption that the color 
electric field is squeezed into flux tubes by the super monopole current. 

%To elucidate the confinement mechanism a little further, we 
%have decomposed the abelian field into its monopole part, 
%which carries the monopole current, and the photon part, which 
%carries the electric current.
%%~\cite{miya}.  
%When two external color electric charges of opposite sign 
%are put into the QCD vacuum, the induced electric field shows up primarily in 
%the photon field, while the monopole super current induces an electric field 
%which exactly cancels the external field \cite{dikk2}.
% As the signal from the monopole part 
%of the field alone is much cleaner than that from the total abelian field, it 
%sometimes helps to consider this part of the distribution only.   

We expect the flux tube to disappear eventually 
if the static charges are
pulled apart beyond a certain distance. 
This distance is expected to be 
around 1.2 fm \cite{schi}. 
%In Fig. 3 we show 
%the distribution of the color electric field in the vicinity of the flux tube
%on the $24^3 48$ lattice 
%at $\beta = 5.29, \kappa=0.1355$, corresponding to
%$m_\pi/m_\rho$ ratio of $ \approx 0.7$, for $R=22$ which amounts to 
We have studied the distribution of the color electric field for 
Q$\bar{\rm Q}$ separation  up to 
$\approx 2$ fm. We observed no sign of string breaking.
This is probably because the abelian Wilson loop has only small overlap
with the broken string as it is the case for the nonabelian
Wilson loop \cite{schi}.

\section{Flux of 3 Quark system} 

For more than 20 years, the question 
whether there is a genuine three body force, 
or the interaction can be described by the sum of the two-body forces
is unanswered \cite{aft,tmns}.
In the former case, the flux structure is expected to be 
of $Y$ shape,
which has a junction at the point where the total length
of strings connecting every quark to the junction is minimal. 
In the latter case, the flux structure
is expected to be of $\Delta$-shape, i.e.  consisting of
three sets of two-body interactions.    

%\begin{figure}[t!b]
%\epsfxsize=8.cm
%\vspace{-4cm}
%\hspace{2cm}
%\includegraphics[width=8cm,angle=270]{E.f420.ps}
%\vspace{-0.5cm}
%\caption{Color flux tube for $R\sim$ 2fm in full QCD.}
%\end{figure}

\begin{figure}[b]
%\vspace*{-0cm}
%\vspace{3cm}
%\begin{flushright}
%\begin{center}
%\vspace{3cm}
\epsfxsize=13.cm
%\vspace{-4cm}
\hspace{.5cm}
%\epsfbox{3QFA455action3.black.eps}
\epsfbox{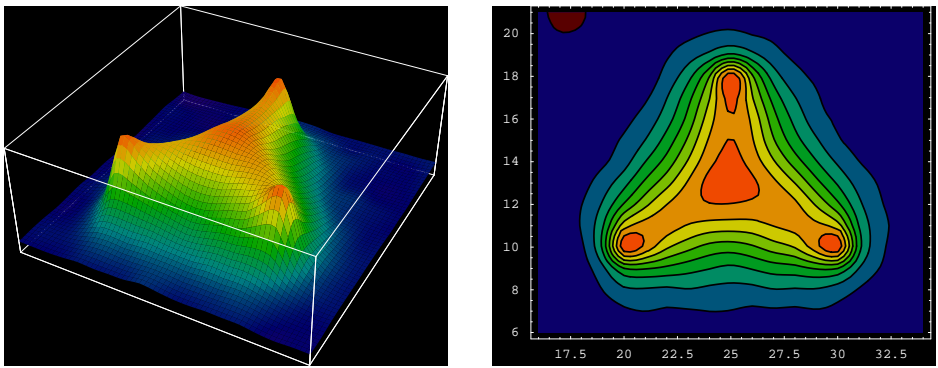}
\vspace{-0.5cm}
%\end{flushright}
\caption{Action density in 3 quark system in full QCD.}
%\end{center}
\end{figure}

Fig. 3 shows the abelian action density of the 3Q system in full QCD.
One can see a clear 
junction in the center of the 3 quarks, which suggests 
 $Y$ shape. 
Fig. 4 shows green component of the color electric field and 
super monopole current  in 3Q system.  
The electric field flows from the green quark to red and 
blue quarks. 
For each leg of the flux, we observe the solenoidal monopole current 
in the plane perpendicular to the flux. 
The strength of the monopole current is proportional to the strength
of the electric flux.
%Corresponding to the amount of flux of the electric field, 
%the monopole current appears strongly around before the junction.   
Our results are in  qualitative 
agreement with the dual Ginzburg-Landau theory \cite{kist,chko}.

In conclusion, we have studied the abelian flux tube in Q$\bar{\rm Q}$
and 3Q systems in full QCD.
The abelian flux tube agrees with the dual superconducting 
scenario, similarly to gluodynamics. 
We made a first study of the abelian flux in the 3-quark system 
in lattice QCD with dynamical quarks.  Our results support  $Y$ shape.

\begin{figure}[t!b]
%\vspace*{-0cm}
%\vspace{3cm}
%\begin{flushright}
%\begin{center}
%\vspace{3cm}
\epsfxsize=11.3cm
\vspace{-0.9cm}
\hspace{0.5cm}
\psfrag{Monopole current}{\small Monopole current}
\psfrag{Electric Field (Green)}{\small Electric field (Green)}
\psfrag{x}{$x$}
\psfrag{y}{$y$}
\psfrag{z}{$z$}
\psfrag{x=21}{{\small $x$=21}}
\psfrag{x=29}{{\small $x$=29}}
\psfrag{y=16}{{\small $y$=16}}
\includegraphics[width=10cm,angle=270]{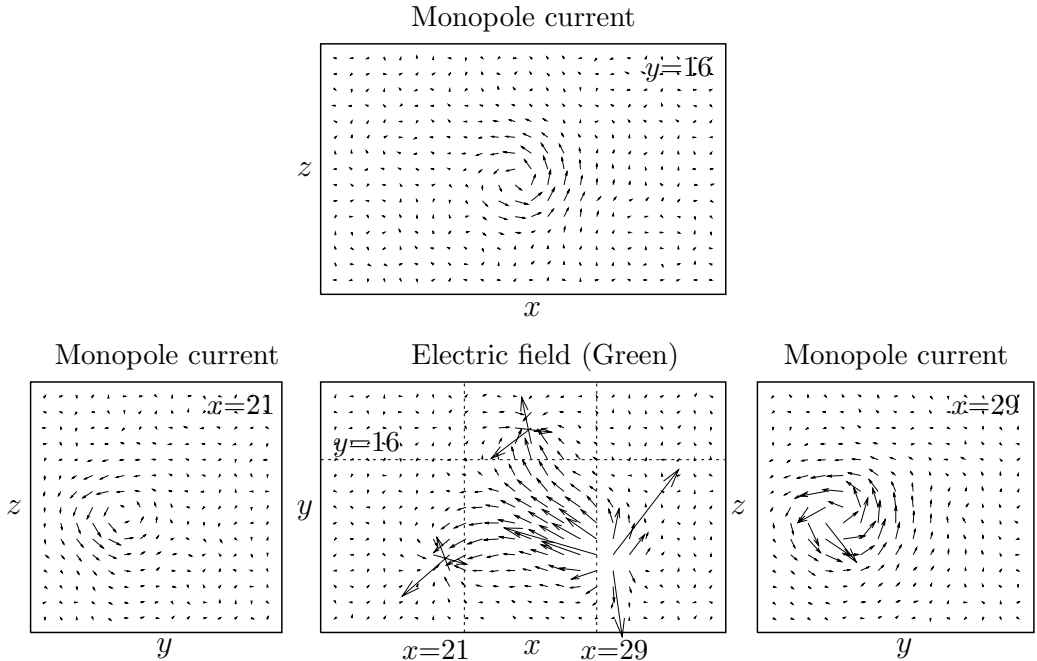}
%\epsfbox{E.f420.ps}
\vspace{-1.5cm}
%\end{flushright}
\caption{Green component of the electric field 
and solenoidal monopole current in 3Q system. The color of the leftmost (rightmost) [top] charge is blue (green) [red]. }
\vspace{-.3cm}
%\end{center}
\end{figure}

\section*{ACKNOWLEDGEMENTS}
%\vspace{-.4cm}
The authors wish to thank T.Suzuki, Y.Mori, Y.Koma, H.Suganuma for  
useful discussions. H. I. thanks the Humboldt University and 
Kanazawa University for hospitality.
The calculations 
have been done on COMPAQ AlphaServer ES40 at Humboldt University.
%\vspace{-.4cm}

\end{document}